\documentclass[letterpaper]{jpconf}
\usepackage{graphicx}
\begin{document}
\title{ARIEL experiments and theory}

\author{Petr Navr\'atil}

\address{TRIUMF, 4004 Wesbrook Mall, Vancouver, British Columbia, V6T 2A3, Canada}

\ead{navratil@triumf.ca}

\begin{abstract}
I present an overview of experiments at TRIUMF ARIEL and ISAC facilities covering both the current and the future envisioned programs. I also briefly review theory program at TRIUMF that relates to the ARIEL experimental program. I highlight several recent experimental results from the nuclear astrophysics, nuclear structure, fundamental symmetries, and the sterile neutrino search. Finally, I mention ongoing theoretical {\it ab initio} calculations of the proton capture on $^7$Li related to the X17 boson observation.
\end{abstract}

\section{Introduction}

\begin{figure}
\begin{center}
\includegraphics[scale=1.0]{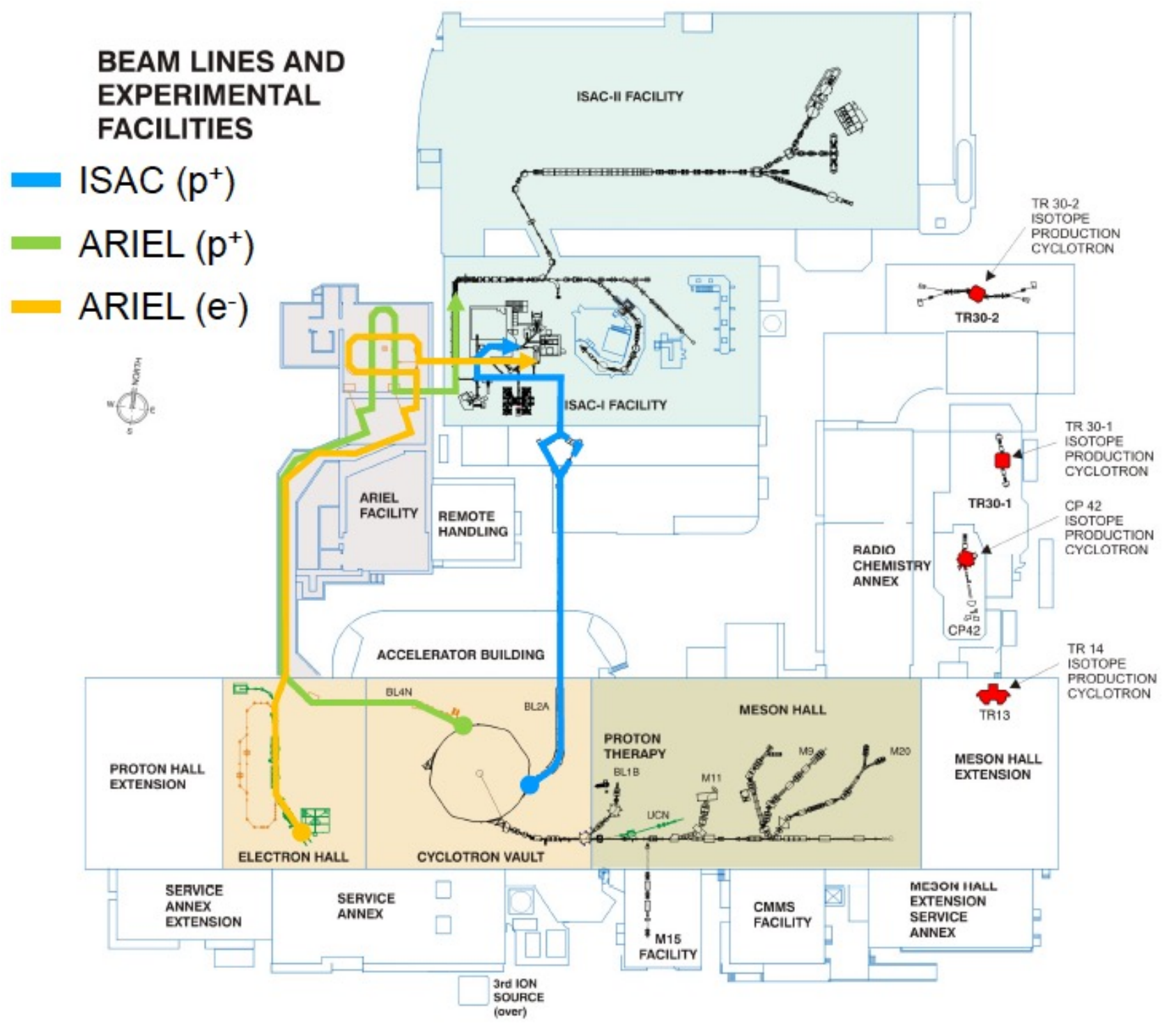}
\end{center}
\caption{\label{fig:TRIUMF_accel}Site view of the TRIUMF accelerator complex.}
\end{figure}

TRIUMF is Canada's particle accelerator centre established in 1968 as the TRI-University Meson Factory. Since 1974, TRIUMF operates the 500 MeV cyclotron accelerating the H$^-$ ions. The high-intensity 500 MeV proton beam drives the radioactive ion beam (RIB) Isotope Separator and ACcelerator (ISAC) facility~\cite{ISAC_Dilling_2014}. The ISAC started its operation in 1995. TRIUMF has recently embarked on the construction of ARIEL, the Advanced Rare Isotope Laboratory, with the goal to significantly expand the RIB program for Nuclear Physics, Life Sciences, and Materials Science~\cite{ARIEL_Dilling_2014}. The nuclear physics program will expand its main pillars, the nuclear structure, astrophysics, and tests of fundamental symmetries. At the heart of ARIEL there is a 100 kW, 30 MeV electron accelerator (e-linac) for isotope production via photo-production and photo-fission as well as a second proton beam line from TRIUMF’s 500 MeV cyclotron for isotope production via proton-induced spallation, fragmentation, and fission. Also included in ARIEL are two production targets and related infrastructure, mass-separators and ion beam transport to ISAC, and a electron-beam ion source (EBIS) for charge breeding. Fig.~\ref{fig:TRIUMF_accel} shows a site view of the TRIUMF accelerator complex. ARIEL will establish a multi-user capability with up to three simultaneous RIBs with more and new isotopes for TRIUMF users. The project completion is planned in 2026 with phased implementation, interleaving science with construction.

\section{ISAC/ARIEL experimental stations}

The short-lived isotopes produced at ISAC and ARIEL targets can be delivered to experiments using a low-energy ($<60$ keV) electrostatic beam transport system or re-accelerated beams. For re-accelerated beams two acceleration stages are available, the first for energies of 0.15 to 1.8 AMeV, designed for high intensity RIBs for studies of nuclear reactions relevant for explosive nucleosynthesis, and a second superconducting stage for energies of at least 6 AMeV for masses of $A<150$.

\begin{figure}
\begin{center}
\includegraphics[scale=0.57]{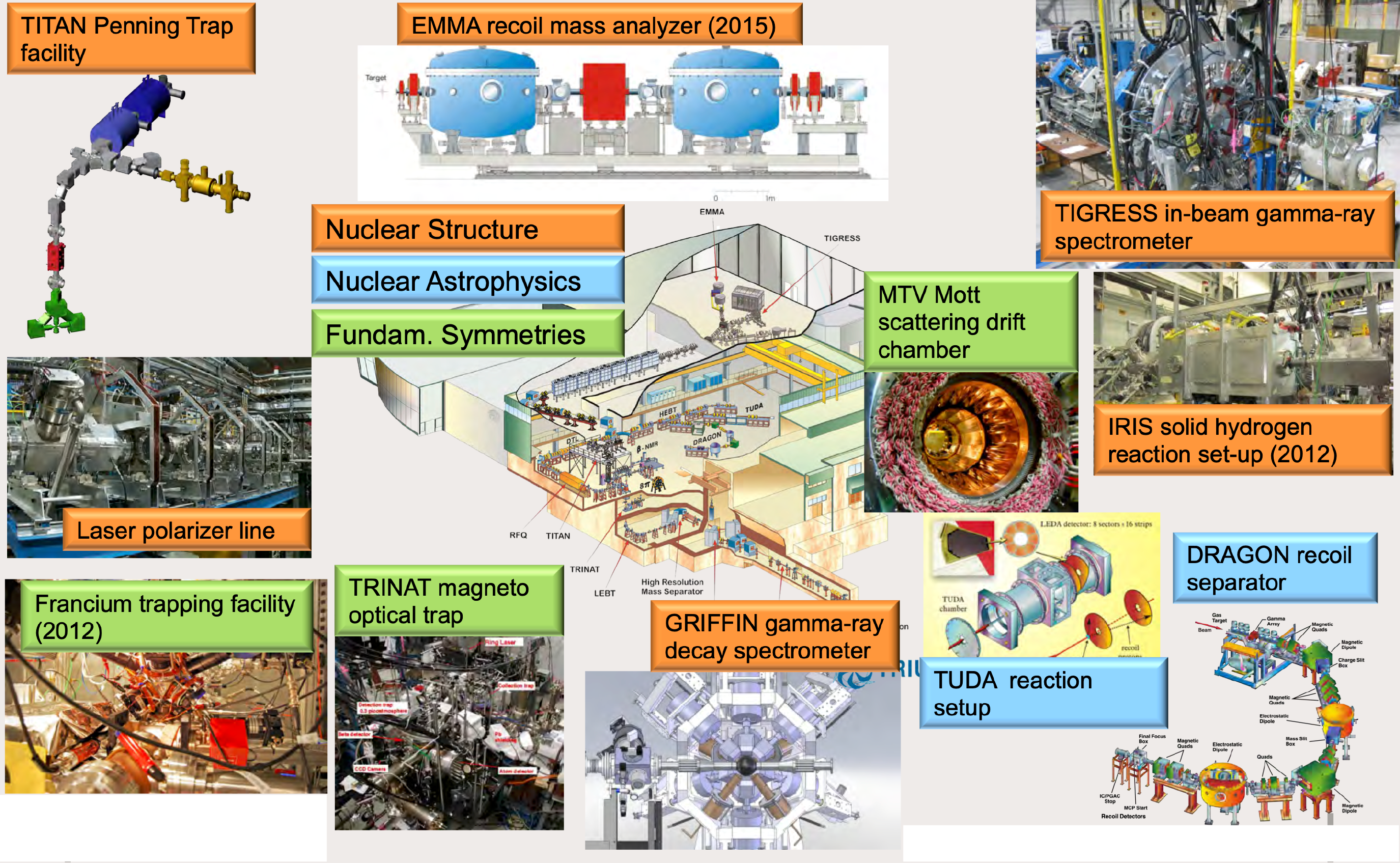}
\end{center}
\caption{\label{fig:ISAC_expt}TRIUMF ISAC/ARIEL experimental stations. Figure courtesy of Jens Dilling.}
\end{figure}

A wide range of experimental stations is available at the ISAC/ARIEL facility to do research in nuclear astrophysics, nuclear structure and reactions, and electro-weak interactions, see Fig.~\ref{fig:ISAC_expt}. In addition to this main focus of the facility, ISAC also hosts world-wide unique $\beta-$NMR and $\beta-$QMR set-ups for the study of magnetic properties in thin materials and at material interfaces.

Instruments focused on nuclear astrophysics research include the DRAGON (Detector of Recoils And Gamma rays Of Nuclear reactions) recoil separator designed to measure radiative capture of RIBs on hydrogen and helium target nuclei, and the TUDA (TRIUMF UK Detector Array) charged particle detector array.

The nuclear structure focused gamma-ray spectroscopy program is centred on the TRIUMF-ISAC $\gamma$-ray escape suppressed spectrometer (TIGRESS), a next generation array of high-efficiency segmented HPGe detectors with digital signal processing that is specifically designed to meet the challenges of experiments with reaccelerated RIBs. A number of auxiliary detectors are used with TIGRESS including a silicon barrel for detecting charged particles SHARC, an array of neutron detectors DESCANT and a recoil mass spectrometer EMMA.

Gamma-Ray Infrastructure For Fundamental Investigations of Nuclei (GRIFFIN) is a state-of-the-art new high-efficiency gamma-ray spectrometer for decay spectroscopy research with the low-energy RIBs. GRIFFIN was installed in 2015 and supports a broad program in nuclear structure, nuclear astrophysics, and fundamental particle interactions.

Nuclear structure and reactions studies of exotic nuclei are also performed with the IRIS (Innovative Rare Isotope reaction Spectroscopy) facility with a frozen (solid) windowless hydrogen or deuterium target and charged particle spectrometer.

A program of high precision mass measurements is carried out at the TITAN (TRIUMF’s Ion Trap for Atomic and Nuclear science) Penning trap facility. Research focus includes nuclear structure studies of exotic halo nuclei~\cite{PhysRevLett.108.052504}, shell evolution in neutron rich isotopes~\cite{PhysRevLett.120.062503,PhysRevC.105.L041301,SILWAL2022137288}, precise determination of $Q$-values of superallowed $\beta$-decays used for the determination of the $V_{ud}$ matrix element of the CKM matrix~\cite{PhysRevC.96.052501}.

Precision measurements of the angular correlations between electron (positron) and anti-neutrino (neutrino) emitted in the $\beta$-decay of spin aligned nuclei allows to search for new components of the electro-weak interaction, such as the exchange of new bosons not described in the Standard Model. Since the neutrino cannot be detected the correlation is determined from the measurement of the momenta of the electron and recoiling nucleus. The TRINAT experiment employs and magneto-optical trap (MOT) for such experiments. Recently, the TRINAT collaboration measured the $\beta$ asymmetry with respect to the initial nuclear spin in $^{37}$K that provided new constraints on physics beyond the standard model~\cite{PhysRevLett.120.062502}.

The Francium trapping facility aims at measurements of the nuclear anapole moment in unstable Fr isotopes that provides a unique access to neutral weak interactions inside the nucleus~\cite{PhysRevA.97.042507}.

\section{Theory program at TRIUMF}

TRIUMF maintains an active theory program focused on {\it ab initio} nuclear theory, nuclear astrophysics, and particle physics. The nuclear theory research is well aligned with the ISAC/ARIEL experiments.

First principles or {\it ab initio} nuclear theory describes nuclei as systems of protons and neutrons interacting by nucleon-nucleon (NN) and three-nucleon (3N) interactions derived from the Quantum Chromodynamics by means of the chiral effective field theory (EFT). With these interactions as input, one solves many-nucleon Schr\"odinger equation to predict properties of atomic nuclei. Unique to TRIUMF nuclear theory is the development of a unified approach to nuclear structure and reactions for light nuclei, the No-Core Shell Model with Continuum (NCSMC) method~\cite{Navratil2016}, and, further, a powerful valence-space method for medium mass nuclei, the Valence-Space In-Medium Similarity Renormalization Group (VS-IMSRG) method~\cite{PhysRevLett.106.222502,PhysRevLett.113.142501}. These approaches involve large-scale high-performance computing using massively parallel codes running on supercomputers including Summit at ORNL and Niagara of the Digital Research Alliance of Canada. Applications of {\it ab initio} nuclear theory include studies of the structure exotic nuclei, nuclei far from stability, nuclear reactions important for astrophysics, tests of fundamental symmetries and searches for physics beyond the standard model. 

Nuclear astrophysics efforts center on theoretical investigations and modelling of the r-process nucleosynthesis. ARIEL will produce unprecedented intensities of rare isotope beams of neutron-rich nuclei through photo-fission that are synthesized in the r-process in the Cosmos. In synergy with the with precision astronomy and sophisticated astrophysical modelling, studies of the neutron rich nuclei will allow us to identify the astrophysical site of the r-process.

Particle theory at TRIUMF focuses on dark matter physics, collider phenomenology, neutrino physics, particle cosmology, and hadronic physics.

\section{Recent research highlights}

This section presents a selection of recent results obtained by ISAC/ARIEL experiments. Synergy with the TRIUMF theory is highlighted.

\subsection{Mass measurements of neutron-rich Titanium and Fe isotopes}

TITAN collaboration performed a precision mass investigation of the neutron-rich titanium isotopes $^{51-55}$Ti~\cite{PhysRevLett.120.062503}. The range of the measurements covered the $N{=}32$ shell closure, and the overall uncertainties of the $^{52-55}$Ti mass values were significantly reduced. Obtained results conclusively established the existence of the weak shell effect at $N{=}32$, narrowing down the abrupt onset of this shell closure. Experimental data were compared with state-of-the-art {\it ab initio} calculations including the in-house VS-IMSRG. In addition, a locally developed chiral EFT 3N interaction was applied and tested for the first time. These calculations, despite very successfully describing where the $N{=}32$ shell gap is strong, overpredicted its strength and extent in titanium and heavier isotones.

Still more recently, high-precision mass measurements of neutron-rich Fe isotopes were performed at the TITAN facility using the newly commissioned multiple-reflection time-of-flight mass spectrometer (MR-ToF-MS)~\cite{PhysRevC.105.L041301}. These measurements were accompanied by in-house {\it ab initio} VS-IMSRG calculations which enabled theoretical assignment of the spin-parities of the $^{69}$Fe ground and isomeric states. Together with mean-field calculations of quadrupole deformation parameters for the Fe isotope chain, these results benchmarked a maximum of deformation in the $N{=}40$ island of inversion in Fe.

\subsection{$^{10}$C elastic scattering with protons and nuclear forces}

\begin{figure}
\begin{center}
\includegraphics[scale=0.4]{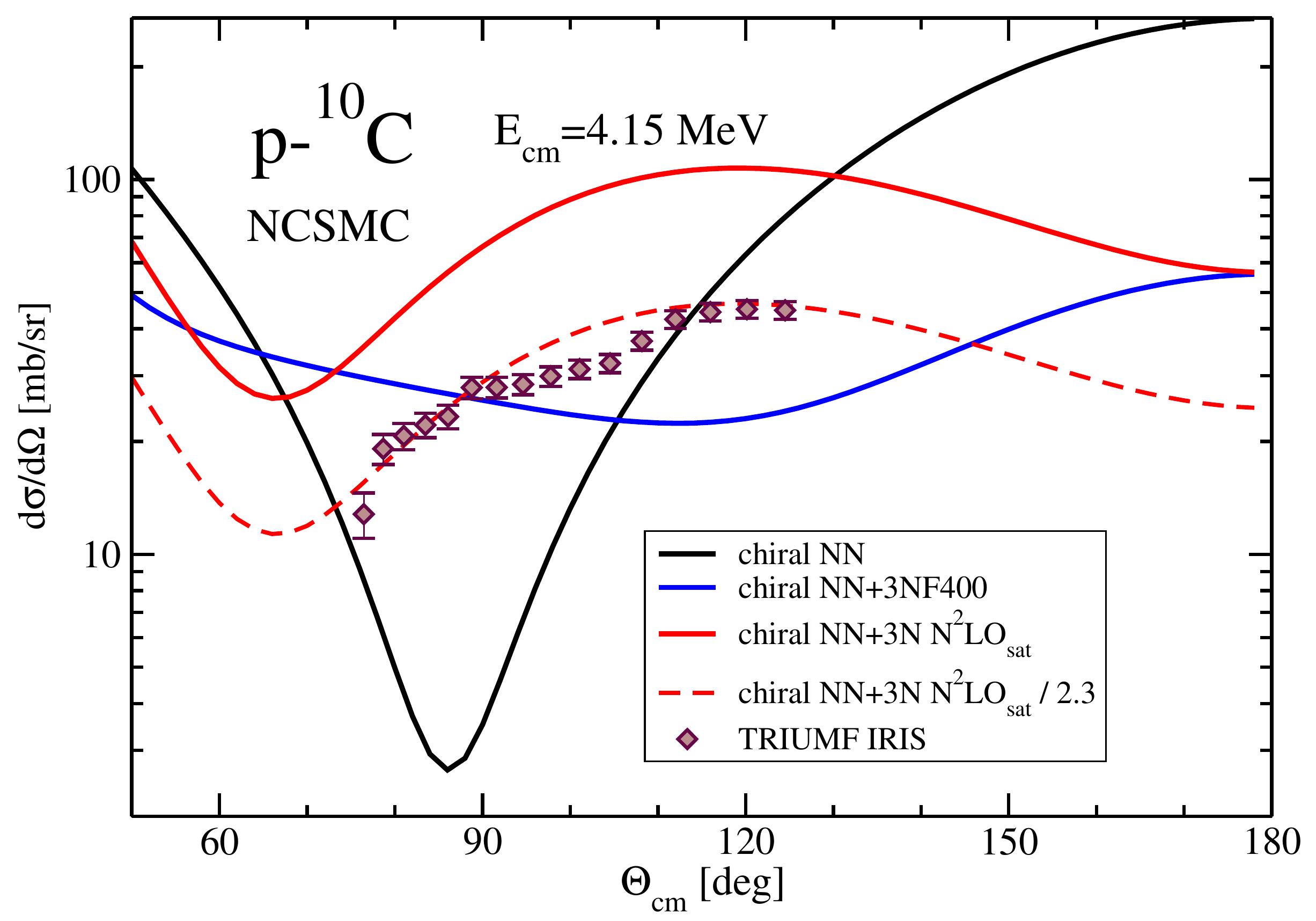}
\end{center}
\caption{\label{fig:pC10}Measured differential cross section for $^{10}$C($p,p$)$^{10}$C$_{gs}$ at $E_{c.m.}{=}4.15$ MeV. The curves are {\it ab initio} theory calculations. The black, blue, and red solid curves are with the chiral NN, NN+3N400, N$^2$LO$_{\rm sat}$ interactions, respectively. The red dashed curve is the N$^2$LO$_{\rm sat}$ calculation scaled down by a factor of 2.3. Further details can be found in Ref.~\cite{PhysRevLett.118.262502}.}
\end{figure}
TRIUMF IRIS reaction spectroscopy station was used to measure elastic scattering of the exotic $^{10}$C nucleus on protons in inverse kinematics~\cite{PhysRevLett.118.262502}. The $^{10}$C nucleus is located at the proton drip line with its isotonic neighbours $^9$B, $^8$Be, and $^{11}$N unbound. The chosen energies covered a low-energy resonance region in the composite unbound $^{11}$N nucleus, a mirror of the halo nucleus $^{11}$Be famous for its parity-inverted ground state. The measured differential cross section was compared to {\it ab initio} NCSMC calculations with different chiral EFT interactions as input, see Fig.~\ref{fig:pC10}. The shape of the cross section was reproduced only with the chiral NN+3N interaction that simultaneously reproduces the $^{11}$Be parity inversion. The experiment demonstrated a strong sensitivity of the elastic scattering off the extremely exotic $^{10}$C to the nuclear force prescription. 

\subsection{Coulomb-excitation measurements of $^{23}$Mg and $^{23}$Na using the TIGRESS spectrometer}

Coulomb-excitation measurements of the $|T_z|{=}1/2$ mirror nuclei $^{23}$Mg and $^{23}$Na, where the former is unstable and the latter stable, were performed at the TRIUMF-ISAC facility using the TIGRESS spectrometer~\cite{PhysRevC.105.034332} . They were used to determine the $E2$ matrix elements of mixed $E2/M1$ transitions between the ground and the first excited state with the goal to challenge {\it ab initio} VS-IMSRG nuclear theory. The experiment improved the precision of the extracted $E2$ strength by a factor of six and demonstrated an underprediction of the strength by the {\it ab initio} theory. This points to the need of including higher-order operator terms to improve the theoretical description.

\subsection{Radioactive capture on nuclear isomers}

The presence of radioactive $^{26g}$Al ($t_{1/2}= 7.2 \times 10^5$ yr) in the universe has been inferred from the detection of characteristic 1.809-MeV $\gamma$ rays throughout the interstellar medium and observations of $^{26}$Mg excesses in meteorites. These discoveries provided direct evidence for ongoing nucleosynthesis in our Galaxy. However, stellar nucleosynthesis of $^{26}$Al is complicated by the presence of a $0^+$ isomer, $^{26m}$Al ($t_{1/2}= 6.3$ s) and, in particular, its destruction by a proton capture reaction needs to be understood. TRIUMF DRAGON spectrometer was used to perform a direct measurement of the $^{26m}$Al($p,\gamma$) reaction~\cite{PhysRevLett.128.042701}.  This is the first direct measurement of an astrophysical reaction using a radioactive beam of isomeric nuclei. The isomeric component of the incoming beam was identified by its associated super-allowed $\beta^+$ decay to the $^{26}$Mg ground state. The strength of a key resonance has been determined which enabled to establish the reaction rate in astrophysical relevant temperature range.

\subsection{The BeEST experiment}

Sterile neutrinos are natural extensions to the standard model of particle physics. A new search for the existence of sub-MeV sterile neutrinos using the decay-momentum reconstruction technique in the decay of $^7$Be has been established by a Colorado School of Mines-TRIUMF-LLNL Beryllium Electron capture in Superconducting Tunnel junctions (BeEST) collaboration~\cite{PhysRevLett.125.032701,PhysRevLett.126.021803,https://doi.org/10.48550/arxiv.2112.02029}. The experiment measures the total energy of the $^7$Li daughter atom from the electron capture decay of $^7$Be implanted into sensitive superconducting tunnel junction (STJ) quantum sensors. The $^7$Be production and implantation is executed at the TRIUMF ISAC/ARIEL facility. Despite its recent launch, the experiment already significantly improved the limits on the existence of sterile neutrinos in the 100–850 keV mass range~\cite{PhysRevLett.126.021803}. TRIUMF particle theorists contribute to the BeEST program.

\section{Future experimental facilities at ARIEL}

Several new facilities as well as enhancements of existing experiments are envisioned. 

\subsection{Delivery of spin-polarized beams to GRIFFIN}

It is planned to deliver spin-polarized beams to the GRIFFIN high-efficiency gamma-ray spectrometer. A new 2-meter section of beamline is needed to join the Polarizer beamline to the GRIFFIN spectrometer. This will allow to perform decay spectroscopy with $\beta-\gamma$ and $\gamma-\gamma$ coincidences of spin-polarized beams including $\gamma$-tagged $\beta$ asymmetry for firm assignment of spins, parities; isospin mixing measurements relevant to the $V_{ud}$ determination; searches for time-reversal breaking. Further, decay spectroscopy of isomerically pure beams by resonant photoionization will become feasible.

\subsection{Radioactive Molecule (RadMol) Laboratory}

Building on TRIUMF's capability to produce a large variety of radioactive ion beams, the international RadMol collaboration seeks to establish a dedicated laboratory for radioactive molecules and fundamental physics at TRIUMF. It will host three independent experimental stations which are directly coupled via a low-energy beamline to TRIUMF's RIB facilities ARIEL and ISAC. The laboratory will also provide space and infrastructure for the ion reaction cell as one formation site for radioactive molecules. The facility plans include a dedicated laser laboratory.

Radioactive molecules can be used as novel precision probes for fundamental physics. Initial physics program will study octuple-deformed nuclei incorporated into polar molecules which provide unmatched sensitivity for nuclear electric dipole moment (EDM). Utilization of diatomic molecules will enable access nuclear anapole moments. Expansions into other fields are also envisioned by the RadMol collaboration.

\subsection{TRIUMF Storage Ring (TRISR)}

\begin{figure}
\begin{center}
\includegraphics[scale=1.0]{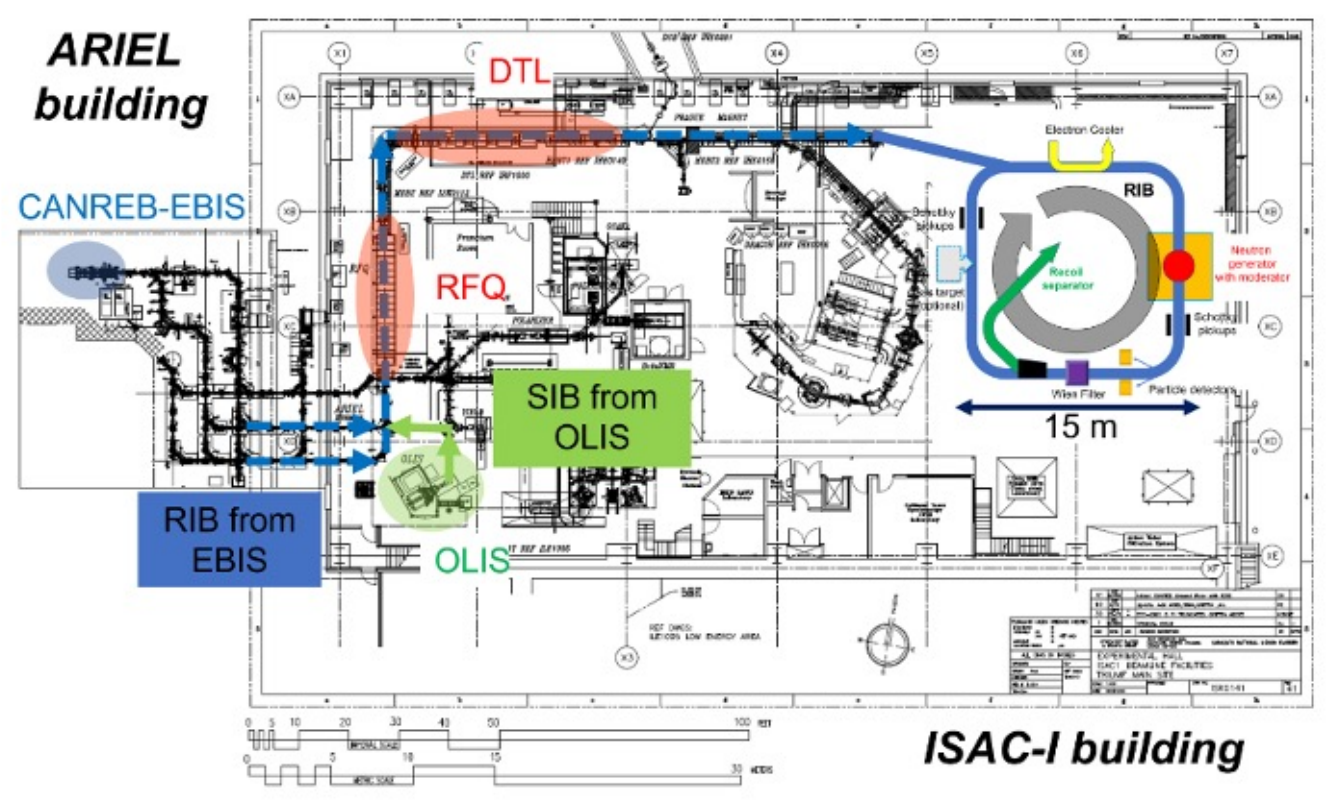}
\end{center}
\caption{\label{fig:TriSR}Proposed TRIUMF Storage Ring for a direct measurement of the neutron capture on radioactive nuclei in inverse kinematics. Figure courtesy of Iris Dillmann.}
\end{figure}
A feasibility study is currently underway for a new worldwide unique facility, the TRIUMF Storage Ring (TRISR) Project for the direct measurement of neutron capture cross sections on short-lived radioactive nuclei down to seconds half-lives in inverse kinematics. The TRISR will be coupled to the ISAC/ARIEL radioactive beam facility and include a high-flux neutron generator producing a “neutron target” that will intersect with the orbiting ion beam. 

Combining the ARIEL e-linac photofission RIB production with the CANREB EBIS will enable cleaner high-intensity neutron-rich beams which are perfectly suited for direct injection into the storage ring. The ion beam orbiting in the storage ring allows multiple passes (100 kHz - few MHz frequency over several seconds) through the target area and thus provides a much higher luminosity than one-pass experiments, enabling a unique physics program. The TRISR will be a low-energy ring (E= $0.1{-}2 \; A$MeV) with a 15m x 15m footprint and will be located next to the DRAGON spectrometer in the ISAC-I experimental hall, see Fig.~\ref{fig:TriSR}. 

The neutron generator and moderator design is presently under discussion. A compact-size high-flux DT neutron generator from the US-based company SHINE Technologies, LLC with an unmoderated neutron flux up to $5\times 10^{13}$ n/s is the favoured candidate. The physics program of the TRISR will center on nuclear astrophysics applications, in particular measurements of radiative neutron capture reaction cross sections on unstable nuclei with $A{>}50$ required for a reliable modelling of the intermediate (i-) and rapid (r-) neutron capture process nucleosynthesis. When no radioactive beam is available, the installation of a diagonal neutron activation position at the neutron generator will in addition allow the production of short-lived radioactive nuclei for use in local radiopharmaceutical research projects (e.g., $^{161}$Tb and $^{177}$Lu for Targeted Radionuclide Therapy and SPECT imaging), as well as for material and environmental science projects to detect trace amounts of stable isotopes via Instrumental Neutron Activation Analysis (INAA).

\section{Theoretical calculations related to the X17 boson}

The DarkLight experiment, a magnetic spectrometer system, will be installed in the ARIEL e-linac beam hall in the next year, with an initial program to search for corroborating evidence of the new physics proposed to explain current anomalies in $^8$Be and $^4$He decay. These anomalies observed in the electron-positron pair production from the proton capture on $^7$Li and $^3$H were reported by the ATOMKI collaboration and interpreted by the decay of a new boson called X17 with the mass ${\sim}17$ MeV~\cite{PhysRevLett.116.042501,Firak2020, PhysRevC.104.044003,https://doi.org/10.48550/arxiv.2205.07744}. The latest developments in the ATOMKI experiments and the progress of the DarkLight installation were reported at the workshop New Scientific Opportunities at the TRIUMF ARIEL e-linac.

\begin{figure}
\begin{center}
\includegraphics[scale=1.0]{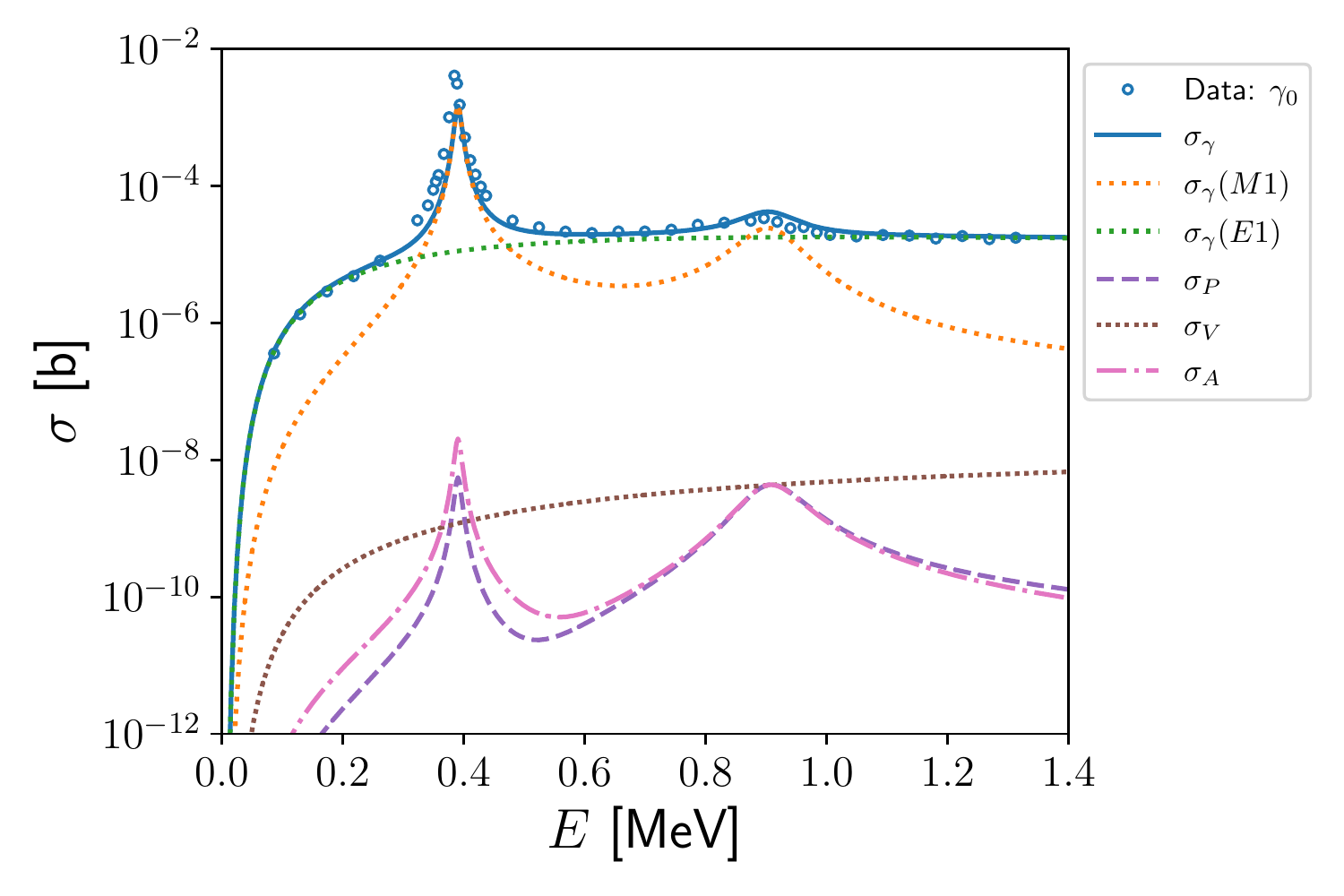}
\end{center}
\caption{\label{fig:Li7p_xsect}The proton capture cross section on $^7$Li dependence on the center-of-mass energy. The calculated $\gamma$ emission cross section is compared to experimental data from Ref.~\cite{Zahnow1995} and to the calculated hypothetical X17 boson emissions considering the $E1$ vector, pseudoscalar (axion), and axial vector boson candidates. All transitions are to the $^8$Be $0^+$ ground state. Calculations are preliminary, details will be given in Ref.~\cite{Gysbers2022}.}
\end{figure}

TRIUMF theorists were active in addressing the X17 boson anomaly shortly after it was initially reported~\cite{PhysRevD.95.115024}. More recently, we embarked on a more in-depth approach to the $p{+}^7$Li capture reaction within the {\it ab initio} NCSMC approach~\cite{Navratil2016}. This way one can describe simultaneously the structure of $^8$Be, the elastic and inelastic $^7$Li($p,p$) scattering, the charge exchange reaction $^7$Li($p,n$)$^7$Be, the $\gamma$ capture $^7$Li($p,\gamma$)$^8$Be, the pair production $^7$Li($p,e^+ e^-$)$^8$Be as well as the X17 boson production $^7$Li($p,X$)$^8$Be and decay for a variety of candidates for the hypothetical boson~\cite{Gysbers2022}. The input for these calculations are chiral EFT NN+3N interactions.  Preliminary NCSMC cross section results are shown in Fig.~\ref{fig:Li7p_xsect}. The calculated $^7$Li($p,\gamma$)$^8$Be radiative capture cross section compares well with the data from Ref.~\cite{Zahnow1995}. The electric dipole ($E1$) and the magnetic dipole ($M1$) contributions are shown separately. The two peaks dominated by the $M1$ contributions are due to the two $1^+$ resonances in $^8$Be with the lower one predominantly isospin $T{=}1$ while the higher one predominantly $T{=}0$ with a suppressed $M1$ decay to the $^8$Be $0^+$ $T{=}0$ ground state. Cross sections for the emission of the X17 boson lower by several orders of magnitude are also shown. Three X17 candidates were considered, a pseudo scalar (axion), axial vector, and $E1$ vector using operators from Refs.~\cite{PhysRevD.95.115024,PhysRevD.18.1607,ZHANG2021136061,PhysRevLett.128.091802}. The electron-positron pair production cross section (not shown in the figure) has a shape basically identical to the $\gamma$ capture cross section with its magnitude scaled by a factor of $\sim \alpha/2\pi \sim 10^{-3}$. One could understand then that no anomaly would be observed in the first resonance with the very high electromagnetic $M1$ rate. An effect from the hypothetical boson can be expected in the second $1^+$ resonance where both the pseudoscalar and the axial vector boson candidate cross sections peak. An anomaly between the $1^+$ resonances would be consistent with the $E1$ vector. The latter is the preferred candidate according to the latest ATOMKI publications~\cite{PhysRevC.104.044003,https://doi.org/10.48550/arxiv.2205.07744}.

\section{Conclusions}
With ARIEL completion within the next five years, TRIUMF will achieve a multi-user capability with up to three simultaneous RIBs. This will allow an expansion of the Nuclear Physics, Life Sciences, and Materials Science program. In synergy with theory, nuclear physics experimental program will be positioned to enhance its science output in all three main pillars, the nuclear structure, astrophysics, and tests of fundamental symmetries. At the same time, ARIEL e-linac will enable particle physics experiments such as DarkLight.

\ack
I would like to thank Iris Dillmann for useful comments. This work was supported by the NSERC Grant No. SAPIN-2022-00019. TRIUMF receives federal funding via a contribution agreement with the National Research Council of Canada. Computing support came from an INCITE Award on the Summit supercomputer of the Oak Ridge Leadership Computing Facility (OLCF) at ORNL, from Livermore Computing, and the Digital Research Alliance of Canada.

\section*{References}


\providecommand{\newblock}{}

\end{document}